# DAN-SNR: A Deep Attentive Network for Social-Aware Next Point-of-Interest Recommendation

Liwei Huang[1], Yutao Ma[2], Yanbo Liu[1], and Keqing He[2]

1. Beijing Institute of Remote Sensing, Beijing 100854, China
E-mails: dr_huanglw@163.com, liuyanbonudt@163.com
2. School of Computer Science, Wuhan University, Wuhan 430072, China
E-mails: ytma@whu.edu.cn, hekeqing@whu.edu.cn

**Abstract:** Next (or successive) point-of-interest (POI) recommendation has attracted increasing attention in recent years. Most of the previous studies attempted to incorporate the spatiotemporal information and sequential patterns of user check-ins into recommendation models to predict the target user's next move. However, none of these approaches utilized the social influence of each user's friends. In this study, we discuss a new topic of next POI recommendation and present a deep attentive network for social-aware next POI recommendation called DAN-SNR. In particular, the DAN-SNR makes use of the self-attention mechanism instead of the architecture of recurrent neural networks to model sequential influence and social influence in a unified manner. Moreover, we design and implement two parallel channels to capture short-term user preference and long-term user preference as well as social influence, respectively. By leveraging multi-head self-attention, the DAN-SNR can model long-range dependencies between any two historical check-ins efficiently and weigh their contributions to the next destination adaptively. Also, we carried out a comprehensive evaluation using large-scale real-world datasets collected from two popular location-based social networks, namely Gowalla and Brightkite. Experimental results indicate that the DAN-SNR outperforms seven competitive baseline approaches regarding recommendation performance and is of high efficiency among six neural-network- and attention-based methods.

## 1. Introduction
### 1.1. Background

Location-based social networks (LBSNs), such as Foursquare[1], Loopt[2], and Yelp[3], have become very popular among young people in the past decade. Users utilize mobile devices and location-based services (LBSs) to search out points of interest (POIs) in LBSNs, post their check-ins and reviews for POIs, and share their life experiences in the real world. Millions of users in LBSNs have generated a massive amount of check-in data, which provides an excellent opportunity to recommend possible POIs for users accurately. Owing to the significance and business value of POI recommendation, the research on POI recommendation has attracted attention from academia and industry. In general, the existing work of POI recommendation attempts to predict target users' preferences based on their historical check-ins and recommend a set of unvisited urban POIs to them [1],[2],[3],[4],[5],[6].

The past few years have witnessed a fast-growing demand for human mobility prediction in urban tourism, product advertising, and other application fields [7]. *Next POI recommendation* has recently emerged as a new research focus of POI recommendation, and its research objective is to predict where target users are likely to go next [8]. To deal with this challenging task, researchers have proposed a few approaches to learn users' movement sequences based on their historical check-ins and train personalized POI recommendation models according to the most recent checked-in locations of users [8],[9],[10],[11]. Due to the complexity and diversity of human mobility, it is difficult for most of the previous studies on next POI recommendation to achieve satisfying recommendation results. Therefore, some recent studies [12],[13],[14],[15] attempted to leverage more available information of user check-ins, such as spatial information and temporal information, to train better recommendation models with deep learning and other new techniques.

---

[1] https://foursquare.com/

[2] http://www.loopt.com/

[3] http://www.yelp.com/

**1.2. Motivation**

As we know, LBSNs are a specific type of online social network that allows users to interact with whomever they like in a virtual world. Intuitively, a user's decision on where to go next may be affected by the user's friends (or called social influence) in LBSNs. Fig. 1 illustrates the concept of social-aware next POI recommendation. Given user1's trajectory of check-ins (sorted in chronological order) at a hotel ($T_{t-5}$), a gym ($T_{t-3}$), and a hotel ($T_{t-1}$), which POIs can be recommended to the user at a given time point ($T_t$)? If a next POI recommendation system does not consider the social influence of the user's friends, it may recommend a restaurant or a museum with equal probability according to the sequential patterns mined from the other four users (or called sequential influence). As shown in Fig. 1, "restaurant" and "museum" appear equally likely to go after "hotel." After leaving a hotel, user2 and user4 like to visit a museum, while user3 and user5 prefer to go to a restaurant. Instead, a social-aware next POI recommendation system, which takes into account user1' friendships with user2 and user4 (see the two red lines between them in Fig. 1), is more likely to recommend a museum to the user at the time point $T_t$.

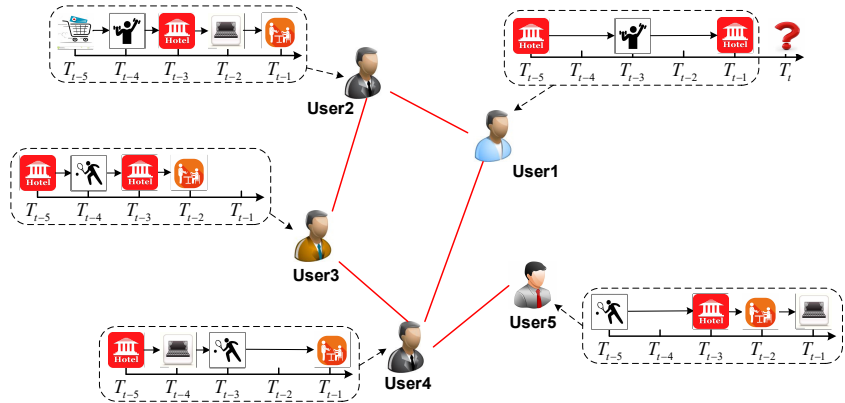

Fig. 1. An example of social-aware next POI recommendation.

A few previous works of POI recommendation have leveraged social influence to improve the quality of recommendations. They calculated the similarities between users regarding friendship and then designed recommendation models using collaborative filtering (CF) techniques [1],[16],[17]. Moreover, some previous studies [18],[19] employed network representation techniques to model users' friendships. However, the above studies showed small improvements in POI recommendation performance because it is tough for them to accurately identify the behavioral correlations between similar users, not just social relationships. In the next POI recommendation scenarios, each user's social influence is dynamic and context-dependent, which causes great difficulty for social-aware next POI recommendation. Therefore, the first problem to solve is the representation of dynamic social influence.

Because the next POI recommendation problem is, in essence, a sequence prediction problem, recurrent neural networks (RNNs) have been recently applied to modeling sequential influence for next POI recommendation [11],[15],[19]. RNNs treat each trajectory as a sequence of user check-ins and recursively compose each check-in behavior with its previous hidden state. Recurrent connections make RNNs applicable to sequence prediction tasks with arbitrary length. However, RNNs also have two disadvantages. Firstly, due to the recursive nature of RNNs, it is hard for them to parallelize [20], which makes both offline training and online prediction very time-consuming. Secondly, fixed-size encoding vectors generated by RNN encoders sometimes do not represent both short and long sequences well [21]. Therefore, the second problem to solve is the efficient and effective joint learning of social influence and sequential influence in the spatiotemporal contexts.

**1.3. Contribution**

For the first problem mentioned above, we attempt to model (dynamic) social influence by capturing the behavioral correlations between each user and the user's friends in LBSNs. Considering the success of attention in natural language processing (NLP), we mine the global dependencies between a user's check-ins and his/her friends' check-ins using the self-attention mechanism [20]. It thus enables our approach to model the context-dependent social influence for next POI recommendation. Like previous studies [18],[19], we also use a graph embedding method to learn network-based user embeddings for each user in a shared latent space.

For the second problem mentioned above, we attempt to find a solution from two aspects. On the one hand, we model sequential influence by capturing the spatiotemporal correlations between each user's check-ins. Due to the disadvantages of RNNs, we also leverage the self-attention mechanism to mine the long-range dependencies between check-ins of a user, which can facilitate the joint learning process of social influence and sequential influence. On the other hand, we use a unified framework of self-attention to model sequential influence and social influence simultaneously. Moreover, we represent each check-in behavior by embedding the spatiotemporal contexts of user check-ins into a compact vector, thus enabling our approach to model user preference better by considering both geographical influence and temporal influence.

In brief, the technical contributions of this work are three-fold.
(1) We first introduce the self-attention mechanism to model dynamic and context-dependent social influence for next POI recommendation. More specifically, we utilize the self-attention mechanism to capture the behavioral correlations between each user's check-ins and his/her friends' check-ins. Also, we learn social-network-based user embeddings by using a graph embedding method.
(2) We present a **d**eep **a**ttentive **n**etwork for **s**ocial-aware **n**ext POI **r**ecommendation (DAN-SNR), which utilizes the self-attention mechanism instead of the architecture of RNNs as a unified framework to model sequential influence and social influence simultaneously. The advantages of the self-attention mechanism can facilitate the parallelization of modeling to speed up the whole joint learning process.
(3) To model user preference better, we take both geographical influence and temporal influence into account and embed the spatiotemporal contexts of user check-ins into a compact vector. More specifically, we construct a location-to-location (L2L) graph based on the distance between POIs to model two-dimensional geographical influence in LBSNs.

To demonstrate the effectiveness of the DAN-SNR, we furthermore evaluated it with two real-world

LBSN datasets, i.e., Gowalla[4] and Brightkite[5]. Experiment results indicate that the DAN-SNR performs better than seven competing baseline approaches of next POI recommendation regarding commonly-used evaluation metrics.

**1.4. Organization**

The remainder of this paper is organized as follows. Section 2 reviews the work related to next POI recommendation and social-aware POI recommendation. Section 3 formulates the problem to solve in this study. Section 4 details the proposed deep attentive network for next POI recommendation. Section 5 presents experiment setups and results. Section 6 discusses some issues related to experimental results. Finally, Section 7 concludes this paper and provides an overview of our future work.

## 2. Related Work
### 2.1. Next POI Recommendation
#### 2.1.1. *Machine-learning-based approach*

As mentioned in Subsection 1.2, the next POI recommendation problem is a problem of sequence prediction, which needs to mine and utilize the sequential influence of users to predict where target users are likely to go next. Many early studies on next POI recommendation employed the Markov chain model to learn sequential influence. For example, Cheng *et al.* [8] designed a matrix factorization model based on a factoring personalized Markov chain model to recommend a successive POI for target users. Zhang *et al.* [22] proposed an additive Markov chain model to predict the sequential probabilities on a location-location transition graph. Ye *et al.* [23] proposed a mixed hidden Markov model to mine the dependencies between POI categories of user check-ins. Because the Markov chain model can model latent check-in behavior patterns, it is still used by a few subsequent studies. For example, Li *et al.* [24] recently proposed a personalized Markov chain model with contextual features (e.g., time of day, day of the week, and POI category) for both the next and next new POI recommendation tasks. However, some recent works on human mobility have revealed that the movement behavior of individuals is not precisely a stochastic process [25], making it hard to meet the underlying assumption of the Markov chain model.

Matrix factorization has been widely used in recommender systems. In addition to the Markov chain model, it is another commonly-used technique to mine sequential patterns of users. For example, Feng *et al.* [9] proposed a personalized ranking metric embedding method to model the user-POI distance and the POI-POI distance in two different hidden spaces, respectively. Zhao *et al.* [10] proposed a spatial-temporal latent ranking method to model the interactions between users and POIs in the fine-grained temporal contexts for successive POI recommendation. Liu *et al.* [26] developed a "Where and When to gO" (WWO) recommender system that recommends possible locations to target users at a specific time point. In particular, the system uses a unified tensor factorization framework to model both static user preference and dynamic sequential influence. He *et al.* [27] also investigated the personalized next POI recommendation problem. They designed a unified tensor-based prediction model that combines the observed sequential patterns and latent behavior preference for each user. However, these approaches built based on matrix factorization often suffer from the cold-start problem.

#### 2.1.2. *Deep-learning-based approach*

Deep learning has recently achieved great success in NLP and computer vision. Some recent works of next POI recommendation began to use RNNs and their variants, such as long short-term memory (LSTM) and gated recurrent unit (GRU), to model sequential influence and temporal dynamics. For example, Liu *et al.* [11] extended the architecture of RNNs and proposed a spatial-temporal recurrent neural network to predict the next location. Due to different levels of sequential patterns in mobile paths, Yang *et al.* [19] employed the RNN and GRU models to represent the short-term and long-term check-in contexts, respectively. Li *et al.* [28] proposed a temporal and multi-level context attention model that is built based on an LSTM-based encoder-decoder framework. The proposed model can learn the spatial-temporal representations of historical check-ins and integrate embedding-based contextual factors in a unified manner. Wu *et al.* [29] designed a long-term and short-term preference learning model. To capture sequential patterns and user preference better, they trained two LSTM networks for location-based sequences and category-based sequences, respectively. Zhao *et al.* [13] proposed a spatiotemporal gated network by improving an LSTM network. They designed spatiotemporal gates that can capture the spatiotemporal relationships between successive check-ins.

---

[4] https://en.wikipedia.org/wiki/Gowalla
[5] https://brightkite.com/

Besides, the attention mechanism [30] has recently been introduced to the architecture of RNNs for next POI recommendation. For example, Huang *et al.* [15] designed an attention-based spatiotemporal LSTM network, which can capture the most critical piece of a user's check-in sequence to predict the next POI. Feng *et al.* [31] developed an attentional recurrent network for mobility prediction from lengthy and sparse user trajectories. Gao *et al.* [32] proposed a variation-attention-based next POI prediction model to overcome the sparsity of user check-ins, with historical mobility attention. Generally speaking, the combination of the attention mechanism with the architecture of RNNs did improve the performance of next POI recommendation based on RNNs. However, the above approaches based on attention and RNNs have the principal disadvantage of high complexity in time.

### 2.2. Social-aware POI recommendation

There is a saying that goes: "Birds of a feather flock together." Inspired by the intuition that friends in LBSNs are more likely to have general preferences and similar behavior patterns, the information of social ties (or relationships) has been leveraged to improve the prediction quality of location-based recommender systems [4],[6]. Previous studies on social-aware POI recommendation usually calculated the similarities between users regarding social relationships (more specifically, friendship) and fused them into the user-based CF approach [3],[5],[16],[17],[33],[34]. For example, Cheng *et al.* [1] further incorporated user similarity into a matrix factorization model. Similarly, Ying *et al.* [35] also incorporated user similarity into a random walk approach. Inspired by the word2vec technique, network representation has become very popular in social networks in recent years. Therefore, some recent studies employed the network embedding method to capture the social influence of friends [18],[19]. However, there is little research that has considered social-aware next POI recommendation. Compared with social-aware POI recommendation, the social influence in the next POI recommendation scenarios is dynamic and context-dependent in LSBNs. Hence, in this study, we need to model the behavioral correlations between each user and his/her friends better.

## 3. Problem Formulation

### 3.1. Notation

Table 1 presents some primary notations used in this paper.

### 3.2. Problem Definition

*Definition 1 (LBSN).* An LBSN is a friend network $G = <U, E>$, where $U$ is a set of users, and $E$ is a set of friendships between users.

*Definition 2 (POI).* In an LSBN, a POI is a spatial item associated with a geographical location, e.g.,

a restaurant or a cinema.

*Definition 3 (Check-in)*. A check-in is a behavior represented by a quintuple $c_t^u = (u, v_t^u, l_t^u, t, p)$, indicating that user $u$ visited POI $v_t^u$ on location $l_t^u$ at time point $t$. Here, $p$ is the position index of $c_t^u$ in the user's trajectory (see Definition 5).

*Definition 4 (Check-in sequence)*. A user's check-in sequence is a set of all the check-ins of the user, denoted by $C_u = \{c_{t_i}^u | 1 \leq i \leq T\}$ where $[t_1, t_T]$ indicates the duration of the user's historical check-ins. For simplicity, historical check-ins of all users are denoted by $C^U = \{C_{u_j} | 1 \leq j \leq |U|\}$.

*Definition 5 (Trajectory)*. A user's trajectory is a set of consecutive check-ins in a session, denoted by $S_{t_k}^u = \{c_{t_{k-M+1}}^u, c_{t_{k-M+2}}^u, \cdots, c_{t_k}^u\}$, where $M$ is the length of the trajectory. $S_{t_k}^u$ is a partially-ordered subset of $C_u$, i.e., $C_u = \bigcup_k S_{t_k}^u$.

Then, we present the problem definition of social-aware next POI recommendation as follows.

*Definition 6 (Social-aware Next POI recommendation)*. Given check-in sequences of all users in an LBSN $G$ at time point $t_N$, the goal of social-aware next POI recommendation is to predict the most likely location $v$ that user $u$ will visit at the next moment $t_{N+1}$, i.e., $\max o_{t_{N+1}, v}^u$.

TABLE 1. Primary notations used in this paper.

| Symbol | Description |
|---|---|
| $u$, $U$ | a user and a set of users |
| $N(u)$ | a set of direct (or one-hop) friends of user $u$ in an LBSN |
| $u' \in N(u) \cup \{u\}$ | an element of the set composed of user $u$ and his/her direct friends |
| $c_t^u = (u, v_t^u, l_t^u, t, p)$ | a check-in: user $u$ visits POI $v_t^u$ on location $l_t^u$ at time point $t$ (position $p$) |
| $C_u$ | a set of check-ins performed by user $u$ |
| $S_{t_k}^u$ | a check-in trajectory of user $u$ |
| $C^U$ | a set of historical check-ins of all users |
| $\mathbf{u}$, $\mathbf{v}_t^u$, $\mathbf{l}_t^u$, $\mathbf{t}$, $\mathbf{p}$ | embeddings of user $u$, POI $v_t^u$, location $l_t^u$, time $t$, and position $p$ |
| $\mathbf{c}_{t_i}^{u,s}$ | the latent representation of check-in $c_{t_i}^u$ generated by the feature embedding layer in the short-term channel |
| $\mathbf{c}_{t_i}^{u,l}$ | the latent representation of $c_{t_i}^u$ generated by the feature embedding layer in the long-term and social channel |
| $\{\mathbf{W}\}$ | a set of parameter matrices in the DAN-SNR |
| $\mathbf{c}_{t_i}^{u,s,k}$, $\mathbf{c}_{t_i}^{u',l,k}$ | the check-in representations generated by the $k$th nonlinear layers in the two channels |
| $\alpha_{ij}^{u,s,k,h}$, $\alpha_{ij}^{u',l,k,h}$ | the attention weights in the $h$th head of the $k$th self-attention layers in the two channels |
| $f(\cdot)$ | the attention function |
| $\mathbf{g}_{t_i}^{u,s,k}$, $\mathbf{g}_{t_i}^{u',l,k}$ | the check-in representations generated by the $k$th self-attention layers in the two channels |
| $\mathbf{c}_v^{u,s}$, $\mathbf{c}_v^{u,l}$ | the latent representations of check-in that user $u$ visits POI $v$ at the next moment in the two channels |
| $\alpha_{iv}^{u,s}$, $\alpha_{iv}^{u',l}$ | the attention weights in the vanilla attention layers in the two channels |
| $\mathbf{h}_{t_N}^{u,s}$, $\mathbf{h}_{t_N}^{u,l}$ | the latent representations of user $u$ generated by the vanilla attention layers in the two channels |
| $o_{t_{N+1}, v}^u$ | the probability that user $u$ visits POI $v$ at the next moment |

## 4. Deep Attentive Network for Social-Aware Next POI Recommendation
### 4.1. Overall Framework

First of all, we assume that each user's check-in behavior is affected by both personal preference and the user's friends. Every user has their own long- and short-term preferences. As for social influence, we consider the behavioral correlations between each user and the user's friends. In this study, we design two parallel channels, namely a short-term channel (STC) and a long-term and social channel (LTSC), to model personal preference and social influence simultaneously. More specifically, the goal of the STC is to learn short-term preference. It takes each user's current trajectory as an input, which represents the user's ongoing sequential influence. Besides, the goal of the LTSC is to learn social influence and long-term preference simultaneously. The LTSC takes all historical check-ins of each user and his/her friends in an LBSN as an input. In particular, we use all historical check-ins previous to the present time point to capture a user's long-term preference.

Because self-attention with time encoding is efficient in both the training and prediction phases, it can be an appropriate replacement for complex RNN structures in sequential behavior encoding [20]. Thus, the two channels of the DAN-SNR also utilize the self-attention mechanism (more specifically, multi-head self-attention) to obtain the representation of each input check-in. In each of the two channels, we perform vanilla attention between the check-in representations and the candidate POI vectors to select valuable check-ins that have more influence on a user's next-step behavior. The final description of the user is then obtained by combining the outputs of the two channels. At last, the DAN-SNR generates the user's preference score for the target POI. As shown in Fig. 2, the overall framework of the DAN-SNR has several building blocks, including a feature embedding layer, $K$ identical nonlinear layers, each of which contains a self-attention layer and a feed-forward layer, a vanilla attention layer, and a prediction component. In the following subsections, we will introduce all the blocks in detail.

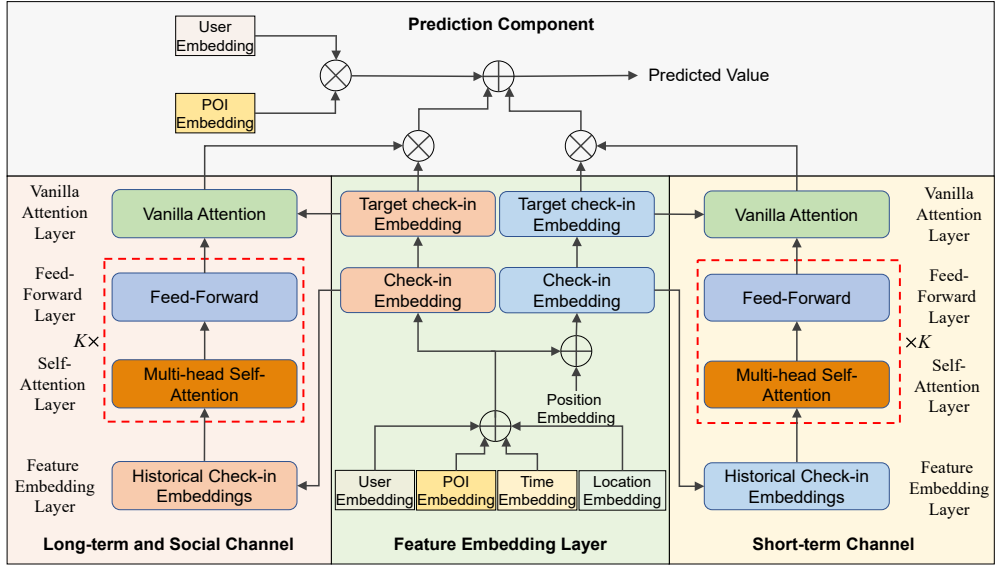

Fig. 2. The overall framework of the DAN-SNR.

### 4.2. Feature Embedding Layer

According to the definition of check-ins, in this study, such behavior has five types of features: user information (more specifically, friendship), POI information, spatial information, temporal information, and position information (in the whole trajectory). We leverage the five types of information, in the form of embedding, to obtain each user's latent representation. Next, we will introduce different embedding techniques used in this study.

#### 4.2.1. *User embedding and POI embedding*

Graph embedding (also known as network embedding) has recently used in many essential tasks on graphs, such as classification and link prediction. Because an LBSN is a friend network in this study, we adopt a graph embedding method, i.e., node2vec [36], to model user information from the viewpoint of friendship. The graph embedding method encodes each node in an LBSN into a low dimensional vector and maintains the structure information of the LBSN. For user $u$, the graph embedding method outputs a feature vector $\mathbf{u}$. By learning the network-based user embedding for each user in a shared latent space, the DAN-SNR can model social relationships for next POI recommendation.

For POI $v_t^u$ in check-in $c_t^u$, we perform a direct lookup on POI IDs and obtain the corresponding POI embedding $\mathbf{v}_t^u$.

#### 4.2.2. *Location embedding*

Previous studies indicate that modeling the geographical influence of user check-ins is essential to POI recommendation in LBSNs. Assuming that strong spatial correlations exist between successive check-ins in short intervals [10], a few researchers attempted to model geographical influence in terms of the geographic distance from users' current locations [8],[9],[10],[11]. However, these studies have some limitations, e.g., they were confined to the one-dimensional geographical influence of locations in a trajectory. Hence, in this study, we attempt to characterize two-dimensional geographical influence via an L2L graph that represents the proximity between POIs regarding geographic distance. An L2L graph is, in essence, a weighted undirected graph, where a vertex represents a POI, a link denotes the spatial correlation between POIs, and the weight of a link indicates geographic distance.

After constructing an L2L graph, we apply the graph embedding method, node2vec [36], on the L2L graph, to encode each location into a low dimensional vector. For location $l_t^u$, node2vec outputs an embedded feature vector $\mathbf{l}_t^u$.

### 4.2.3. *Time embedding*

Human mobility is affected by circadian rhythms, habits and customs, and other factors [37]. The next POI recommendation task is thus time-dependent. In other words, temporal information is critical to analyzing individual check-in behavior. However, it is difficult to learn a proper embedding directly from the continuous-time nature of check-ins using embedding concatenation or addition [21]. In this study, we use a temporal encoding method proposed by Zhou *et al.* [21]. Firstly, we slice the elapsed time, w.r.t the ranking time, into intervals whose length grows exponentially. For example, we can map the time in the range [0,1), [1,2), [2,4), ..., [$2^k$,$2^{k+1}$) to a categorial feature of 0, 1, 2, ..., $k$+1. Different groups of check-in behavior may have different granularities of time slicing. Secondly, we perform a direct lookup on the categorial time features and obtain the time embedding $\mathbf{t}$ of time point $t$.

### 4.2.4. *Position embedding*

As shown in Fig. 2, the DAN-SNR does not contain any recurrence or convolution. To better model sequential influence, we use the timing signal approach proposed by Vaswani *et al.* [20]. Compared with those traditional positional encoding methods, the timing signal approach does not introduce additional parameters. The components of the position embedding $\mathbf{p}$ of postion $p$ can be formulated as follows:

$$\mathbf{p}_{(2i)} = \sin(p/10000^{2i/d}), \tag{1}$$

$$\mathbf{p}_{(2i+1)} = \cos(p/10000^{2i/d}), \tag{2}$$

where $2i$ and $2i+1$ are even and odd integers, respectively, and $d$ is the dimension of latent variables.

### 4.2.5. *Concatenation of different embeddings*

After calculating the embeddings of users, POIs, locations, time, and positions, we then carry out a concatenation operation on them to obtain the hidden representation of each check-in. When modeling short-term user preference, sequential check-in patterns play an essential role in predicting the target user's next POI. Thus, we consider the position embedding in the STC. Instead, the position embedding is not used in the LTSC. It is because we take into account all the historical check-ins of each user and his/her friends. In other words, we believe that short-term sequential patterns have a limited impact on a user's long-term preference and social influence.

For user $u$, we feed feature vectors $\mathbf{u}$, $\mathbf{v}_t^u$, $\mathbf{l}_t^u$, $\mathbf{t}$, and $\mathbf{p}$ into a $d$-dimensional fully-connected layer, which outputs the latent representation of each check-in $\mathbf{c}_t^{u,s}$ (see Eq. (3)) in the STC and $\mathbf{c}_t^{u,l}$ (see Eq. (4)) in the LTSC.

$$\mathbf{c}_t^{u,s} = \text{sigmoid}(\mathbf{W}_u\mathbf{u} + \mathbf{W}_v\mathbf{v}_t^u + \mathbf{W}_l\mathbf{l}_t^u + \mathbf{W}_t\mathbf{t} + \mathbf{W}_p\mathbf{p}), \tag{3}$$

$$\mathbf{c}_t^{u,l} = \text{sigmoid}(\mathbf{W}_u\mathbf{u} + \mathbf{W}_v\mathbf{v}_t^u + \mathbf{W}_l\mathbf{l}_t^u + \mathbf{W}_t\mathbf{t}), \quad (4)$$

where $\mathbf{W}_u \in \mathbb{R}^{d\times d}$, $\mathbf{W}_v \in \mathbb{R}^{d\times d}$, $\mathbf{W}_l \in \mathbb{R}^{d\times d}$, $\mathbf{W}_t \in \mathbb{R}^{d\times d}$, and $\mathbf{W}_p \in \mathbb{R}^{d\times d}$ are transition matrices. Note that the two channels of the DAN-SNR share the same parameters of the concatenation operation.

### 4.3. Self-Attention Layer

As mentioned in the Introduction Section, modeling various types of check-in information in a unified manner is quite challenging for next POI recommendation. For example, check-in behavior has three main temporal properties, namely periodicity, non-uniformness, and consecutiveness [4]. Location information of check-ins also has three unique features, namely hierarchical data, measurable distance, and sequential ordering [6]. Besides, modeling spatiotemporal characteristics of check-ins often relies on different prior assumptions. For example, the spatial distribution of a user's visited locations follows a specific distribution (e.g., the power law), and a user's check-ins are periodic in one day or one week. To address the above problems, in this study, the DAN-SNR uses the self-attention mechanism to model the interactions between user check-ins regardless of what type of check-in information is involved, which can capture social, sequential, temporal, and spatial influence in a unified way. Moreover, the self-attention mechanism can measure the degree of behavioral correlations between check-ins automatically and then adjust the attention weights accordingly to predict the next POI. Thus, the DAN-SNR can work without any prior assumptions.

As shown in Fig. 2, the primary goal of the self-attention layer is to capture two types of behavioral correlations between check-ins. In the STC, the self-attention layer's outputs stand for the representative sequence of check-ins that considers the impact of each user's short-term preference. In the LTSC, the outputs of the self-attention layer indicate the representative series of check-in behavior that takes into account each user's long-term preference and his/her friends' actions. In this work, we leverage a similar multi-head self-attention structure proposed by Vaswani *et al.* [20] for the machine translation task [38], with some customized settings. In theory, we can utilize the multi-head self-attention to calculate the behavioral correlations between check-ins of all users. However, this will cause a tremendous amount of computation. Therefore, we model social influence by capturing only the behavioral correlations between check-ins of direct (or one-hop) friends in an LBSN.

#### 4.3.1. *Modeling short-term preference*

Given the current trajectory of user $u$ $S_{t_N}^u = \{c_{t_{N-M+1}}^{u,s}, c_{t_{N-M+2}}^{u,s}, \ldots, c_{t_N}^{u,s}\}$ in the STC, let $\mathbf{C}_{t_N}^{u,s,k} = [\mathbf{c}_{t_{N-M+1}}^{u,s,k}, \mathbf{c}_{t_{N-M+2}}^{u,s,k}, \ldots, \mathbf{c}_{t_N}^{u,s,k}] \in \mathbb{R}^{d\times M}$ be a matrix that consists of latent feature vectors generated by the $k$th nonlinear layer in the STC. Here, $\mathbf{C}_{t_N}^{u,s}$ represents a matrix that consists of hidden feature vectors generated by the feature embedding layer, $d$ is the dimension of hidden variables, and $M$ is the length of the user trajectory. We can utilize the multi-head self-attention to generate a new representation of check-ins in the user trajectory, described below.

$$\mathbf{r}_{t_i}^{u,s,k} = \text{concat}(\text{head}_1^{u,s,k,t_i}, \text{head}_2^{u,s,k,t_i}, \ldots, \text{head}_H^{u,s,k,t_i})\mathbf{W}^{s,k}, \quad (5)$$

$$\text{head}_h^{u,s,k,t_i} = \sum_{j=N-M+1}^{N} \alpha_{ij}^{u,s,k,h} \mathbf{c}_{t_j}^{u,s,k-1,h}, \quad (6)$$

where $\text{concat}(\cdot)$ denotes a concatenation operation, $H$ is the number of heads in the multi-head self-attention, $\mathbf{W}^{s,k} \in \mathbb{R}^{d\times d}$ is a parameter matrix, $\mathbf{c}_{t_j}^{u,s,k-1,h}$ is the hidden feature vector of the $h$th head divided from $\mathbf{c}_{t_j}^{u,s,k-1}$, and $\alpha_{ij}^{u,s,k,h}$ is the weight of attention.

Then, we will introduce the calculation process of the attention weight matrix $\mathbf{A}^{u,s,k,h} = (\alpha_{ij}^{u,s,k,h})$ in detail. For each pair of latent feature vectors of the input to the $k$th nonlinear layer in the STC, i.e., $(\mathbf{c}_{t_i}^{u,s,k-1,h}, \mathbf{c}_{t_j}^{u,s,k-1,h})$, the attention weight $\alpha_{ij}^{u,s,k,h}$ measures the degree of the former's impact on the latter. More specifically, we calculate this parameter using the following equation:

$$\alpha_{ij}^{u,s,k,h} = \frac{\exp(f(\mathbf{c}_{t_i}^{u,s,k-1,h}, \mathbf{c}_{t_j}^{u,s,k-1,h}))}{\sum_{j=N-M+1}^{N} \exp(f(\mathbf{c}_{t_i}^{u,s,k-1,h}, \mathbf{c}_{t_j}^{u,s,k-1,h}))}, \tag{7}$$

where $f(\mathbf{c}_{t_i}^{u,s,k-1,h}, \mathbf{c}_{t_j}^{u,s,k-1,h})$ is an attention function. As mentioned above, we use the dot-product attention as the attention function in this study. It is because Vaswani *et al.* [20] found that the additive attention outperforms the dot-product attention when $d$ is large. As with the work [20], we also define the attention function with a scale, described as follows.

$$f(\mathbf{c}_{t_i}^{u,s,k-1,h}, \mathbf{c}_{t_j}^{u,s,k-1,h}) = \frac{\mathbf{c}_{t_i}^{u,s,k-1,h}(\mathbf{c}_{t_j}^{u,s,k-1,h})^T}{\sqrt{d}}. \tag{8}$$

4.3.2. *Modeling long-term preference and social influence*

Given user $u$ and the user's direct friends $N(u)$ in the LTSC, we consider all historical check-ins of $u$ and his/her friends at time point $t_N$, i.e., $L_{t_N}^u = \bigcup_{t_1}^{t_N} \bigcup_{u' \in N(u) \cup \{u\}} \{c_{t_i}^{u'}\}$. By using the multi-head self-attention, we can calculate the attention weight $\alpha_{ij}^{u',l,k,h}$ for each pair of check-ins $c_{t_i}^{u'}, c_{t_j}^{u'} \in L_{t_N}^u$ in the LTSC, according to Eq. (9) and Eq. (10).

$$\alpha_{ij}^{u',l,k,h} = \frac{\exp(f(\mathbf{c}_{t_i}^{u',l,k-1,h}, \mathbf{c}_{t_j}^{u',l,k-1,h}))}{\sum_{j=1}^{N} \exp(f(\mathbf{c}_{t_i}^{u',l,k-1,h}, \mathbf{c}_{t_j}^{u',l,k-1,h}))}, \tag{9}$$

$$f(\mathbf{c}_{t_i}^{u',l,k-1,h}, \mathbf{c}_{t_j}^{u',l,k-1,h}) = \frac{\mathbf{c}_{t_i}^{u',l,k-1,h}(\mathbf{c}_{t_j}^{u',l,k-1,h})^T}{\sqrt{d}}. \tag{10}$$

Then, we can generate a new vector representation $\mathbf{r}_{t_i}^{u',l,k}$ of check-in $c_{t_i}^{u'} \in L_{t_N}^u$ using the $k$th self-attention layer in the LTSC, described below.

$$\mathbf{r}_{t_i}^{u',l,k} = \text{concat}(\text{head}_1^{u',l,k,t_i}, \text{head}_2^{u',l,k,t_i}, \dots, \text{head}_H^{u',l,k,t_i})\mathbf{W}^{l,k}, \tag{11}$$

$$\text{head}_h^{u',l,k,t_i} = \sum_{j=1}^{N} \alpha_{ij}^{u',l,k,h} \mathbf{c}_{t_j}^{u',l,k-1,h}, \tag{12}$$

where $\mathbf{c}_{t_j}^{u',l,k-1}$ is the output of the $(k-1)$th nonlinear layer, $\mathbf{c}_{t_j}^{u',l,k-1,h}$ is the hidden feature vector of the $h$th head divided from $\mathbf{c}_{t_i}^{u',l,k-1}$, and $\mathbf{W}^{l,k} \in \mathbb{R}^{d \times d}$ is a parameter matrix.

4.3.3. *Residual connection*

He *et al.* [39] indicated that the depth of representations is essential to achieve excellent performance in visual recognition tasks. Inspired by their idea, we construct our model with residual learning [39], which has been demonstrated to be very useful for training deep neural networks. In both the two channels, we add a residual connection to each self-attention layer. Following the work of Vaswani *et al.*

[20], we then apply layer normalization [40] after the residual connection to stabilize the activations of deep neural networks. Given an input $c_{t_i}^{u,s,k-1}$ in the STC, the output $g_{t_i}^{u,s,k}$ of the $k$th self-attention layer in this channel is computed by the following equation:

$$g_{t_i}^{u,s,k} = \text{layer\_norm}(c_{t_i}^{u,s,k-1} + r_{t_i}^{u,s,k}), \tag{13}$$

where layer_norm($\cdot$) denotes the layer normalization function. Note that we can obtain $g_{t_i}^{u',l,k}$ in a similar form of Eq. (13).

### 4.4. Feed-Forward Layer

The learning capability of neural networks depends on highly flexible nonlinear transformations. Unlike neural networks, the self-attention mechanism encodes an input sequence into an output sequence using weighted sum operations. Thus, it has limited capability to represent latent features. To improve the representation capability of the DAN-SNR, we employ a fully connected feed-forward network to deal with the output from each of the self-attention layers. Each feed-forward layer of the DAN-SNR consists of two transformations and a rectified linear unit (ReLU) activation between them. For example, given an input $g_{t_i}^{u,s,k}$ generated by the $k$th self-attention layer in the STC, we can calculate the output $c_{t_i}^{u,s,k}$ of the feed-forward layer using the following equation:

$$c_{t_i}^{u,s,k} = \text{relu}(g_{t_i}^{u,s,k} W_1^s) W_2^s, \tag{14}$$

where relu($\cdot$) denotes the ReLU activation function, and $W_1^s, W_2^s \in \mathbb{R}^{d \times d}$ are two trainable parameter matrices. The linear transformations have the same parameters across different check-in representations in one feed-forward layer, but they vary from layer to layer. Similarly, we can obtain the output of the $k$th feed-forward layer in the LTSC, i.e., $c_{t_i}^{u',l,k}$.

### 4.5. Vanilla Attention Layer

By using the $K$ nonlinear layers, we obtain an abstract representation of each check-in behavior, which encodes social influence, sequential influence, spatial influence, and temporal influence simultaneously. Next, we will leverage these check-in representations to predict the target user's next POI. Because not all the historical check-ins of a user and his/her friends have the same effect on the user's next-step behavior, we need to pay attention to those more useful ones. Therefore, we design a vanilla attention layer to capture the correlations between the representations of historical check-ins regarding the next-step movement. By using the attention mechanism, the vanilla attention layer can help to select the representative check-ins that characterize user preference and social influence, as well as to assign different weights to them in a flexible, efficient way.

We take the vanilla attention layer in the LTSC as an example. Given user $u$, his/her direct friends $N(u)$, and their check-in history before the time point $t_N$, the output of each check-in representation generated by the last feed-forward layer is $c_{t_i}^{u',l,K}$, where $u' \in N(u) \cup \{u\}$ and $t_i \in [t_1, t_N]$. For each candidate POI $v$, we concatenate different embeddings and then employ an attention network and the softmax function to calculate the normalized attention weight, which follows a similar procedure in the self-attention layers.

$$c_v^{u,l} = \text{sigmoid}(W_u u + W_v v_{t_{N+1}}^u + W_l l_{t_{N+1}}^u + W_t t_{N+1}), \tag{15}$$

$$\alpha_{iv}^{u',l} = \text{softmax}(f(c_{t_i}^{u',l,K}, c_v^{u,l})), \tag{16}$$

$$f(\mathbf{c}_{t_i}^{u',l,K}, \mathbf{c}_v^{u,l}) = \frac{\mathbf{c}_{t_i}^{u',l,K}(\mathbf{c}_v^{u,l})^T}{\sqrt{d}}, \tag{17}$$

where $\mathbf{c}_v^{u,l}$ is the latent representation of the check-in behavior that $u$ visits $v$ at the next moment $t_{N+1}$ in the LTSC, and $\alpha_{ij}^{u,l}$ is the attention weight for each pair of $\mathbf{c}_{t_i}^{u',l,K}$ and $\mathbf{c}_v^{u,l}$. Once the vanilla attention layer outputs the attention weights, the hidden representation of $u$ concerning $v$ is calculated using the following equation:

$$\mathbf{h}_{t_N}^{u,l} = \frac{1}{N}\sum_{i=1}^{N} \alpha_{iv}^{u',l} \mathbf{c}_{t_i}^{u',l,K}. \tag{18}$$

For the vanilla attention layer in the STC, we can obtain the hidden representation of $u$ concerning $v$, i.e., $\mathbf{h}_{t_N}^{u,s}$, which is calculated by the following operation:

$$\mathbf{h}_{t_N}^{u,s} = \frac{1}{M}\sum_{i=N-M+1}^{N} \alpha_{iv}^{u,s} \mathbf{c}_{t_i}^{u,s,K}, \tag{19}$$

$$\alpha_{iv}^{u,s} = \text{softmax}(f(\mathbf{c}_{t_i}^{u,s,K}, \mathbf{c}_v^{u,s})), \tag{20}$$

where $\mathbf{c}_v^{u,s}$ is the latent representation of the check-in behavior that $u$ visits $v$ at a specific time point $t_{N+1}$ in the STC. Compared with $\mathbf{c}_v^{u,l}$, $\mathbf{c}_v^{u,s}$ also considers the position embedding in addition to the other four embeddings.

$$\mathbf{c}_v^{u,s} = \text{sigmoid}(\mathbf{W}_u \mathbf{u} + \mathbf{W}_v \mathbf{v}_{t_{N+1}}^u + \mathbf{W}_l \mathbf{l}_{t_{N+1}}^u + \mathbf{W}_t \mathbf{t}_{N+1} + \mathbf{W}_p \mathbf{p}_{N+1}). \tag{21}$$

**4.6. Prediction Component**

In this study, a user's preference score is defined as a function of three embeddings, namely $\mathbf{h}_{t_N}^{u,s}$, $\mathbf{h}_{t_N}^{u,l}$, and $\mathbf{u}$. We recommend possible POIs for the target user by calculating the dot-product of user and POI representations, which is similar to those previous studies using matrix factorization. Finally, the predicted probability that user $u$ visits candidate POI $v$ at time point $t_{N+1}$ (i.e., the preference score) can be obtained by the following equation:

$$o_{t_{N+1},v}^u = \left(\mathbf{h}_{t_N}^{u,s}\right)^T \mathbf{c}_v^{u,s} + \left(\mathbf{h}_{t_N}^{u,l}\right)^T \mathbf{c}_v^{u,l} + (\mathbf{u})^T \mathbf{v}, \tag{22}$$

where the last item of the equation denotes the inherent interest of user $u$ in POI $v$.

**4.7. Model Training**

For user $u$, we build a training instance $I_N = <S_{t_N}^u, L_{t_N}^u, \{(v,v')\}>$ at time point $t_N$, including the current trajectory $S_{t_N}^u$, a set of historical check-ins of the user and his/her friends $L_{t_N}^u$, and all pairs of positive and negative POIs $\{(v,v')\}$ at the next moment $t_{N+1}$. Here, $v$ and $v'$ represent a positive (or called observed) POI and a negative (or called unobserved) POI, respectively. For the construction process of training instances, please refer to *Algorithm 1*.

We use the Bayesian personalized ranking (BPR) [41] rather than the point-wise loss to define loss function for model parameter learning. By learning a pair-wise ranking loss in the training process of the DAN-SNR, BPR can make use of the unobserved user-POI data. Moreover, BPR considers the relative order of POIs to predict users' preference scores, which is based on an underlying assumption that each user prefers the observed POI (or called positive example) over the unobserved POIs (or called negative examples).

The BPR loss function requires pairs of two scores: one for the target POI (i.e., the actual next POI) and the other for a negative sample (i.e., any POI except the target POI). However, calculating scores for all pairs of $(v,v')$ is not practical in real-world application scenarios with millions of items [42]. Thus, we use a sampling method, which is similar to the negative sampling mechanism used in word2vec [43], to select a fraction of POIs as negative samples during the training process. Because POI's geographical information has an essential impact on predicting a user's next movement, we choose negative samples from the observed POIs located in the same city randomly. If the number of all the observed POIs situated in the same town are smaller than the size of negative samples, we employ the popularity-based sampling method [42] to generate the remaining negative samples.

---
**Algorithm 1**: Constructing training instances
**Input**: an LBSN $G$ and a set of historical check-in sequences of all users $C^U$
**Output**: a set of training instances $D$

01. Initialize $D = \bigcup_u D^u = \emptyset$;
02. **For** each user $u$ in $G$ **do**
03.     **For** each user trajectory $S^u_{t_{N+1}}$ in $C_u$ **do**
04.         Get the set of historical check-ins of $u$ and his/her friends at $t_N$, $L^u_{t_N} = \bigcup^{t_N}_{t_1} \bigcup_{u' \in N(u) \cup \{u\}} \{c^{u'}_{t_i}\}$;
05.         Get a positive sample $v^u_{t_{N+1}}$ that $u$ visited at $t_{N+1}$ from $c^u_{t_{N+1}}$;
06.         Get the set of negative samples $\{v'^u_{t_{N+1}}\}$ by the sampling method;
07.         Add a training instance $<S^u_{t_N}, L^u_{t_N}, \{(v^u_{t_{N+1}}, v'^u_{t_{N+1}})\}>$ to $D^u$;
08.     **End for**
09. **End for**
10. $D = \bigcup_u D^u$;
11. **Return** the training set $D$;

---

Then, we use the maximum a posterior (MAP) estimation to learn the parameters of the DAN-SNR, described below.

$$p(u, t_{N+1}, v > v') = g(o^u_{t_{N+1},v} - o^u_{t_{N+1},v'}), \tag{23}$$

where $g(\cdot)$ denotes a nonlinear function defined as

$$g(x) = \frac{1}{1+e^{-x}}. \tag{24}$$

By integrating the pair-wise loss function and a regularization term, we can solve the objective function of the DAN-SNR for the next POI recommendation task as follows.

$$\begin{aligned} J &= -\sum_{I_N} \sum_u \sum_{(v,v')} \ln p(u, t_{N+1}, v > v') + \frac{\lambda}{2} \|\Theta\|^2 \\ &= \sum_{I_N} \sum_u \sum_{(v,v')} \ln(1 + e^{-(o^u_{t_{N+1},v} - o^u_{t_{N+1},v'})}) + \frac{\lambda}{2} \|\Theta\|^2. \end{aligned} \tag{25}$$

where $\lambda$ determines the power of regularization, and $\Theta$ indicates the parameter set. Note that the objective function is optimized by the stochastic optimization method, Adam [44]. For the whole training process of the DAN-SNR, please refer to *Algorithm 2*.

---
**Algorithm 2**: Training the DAN-SNR
**Input**: a training set $D$
**Output**: the parameter set of the DAN-SNR $\Theta$

01. Initialize the parameter set $\Theta$;
02. **While** (exceed(maximum number of iterations) == FALSE) **do**
03.     Randomly select a batch of training instances $D_b$ from $D$;
04.     **For** each user $u$ in $D_b$ **do**
05.         **For** each pair of $(v, v')$ of $u$ in $D_b$ **do**
06.             Calculate the probabilities $o^u_{t_{N+1},v}$ and $o^u_{t_{N+1},v'}$ according to Eq. (22);
07.         **End for**
08.     **End for**
09.     Find $\Theta$ minimizing the objective function (Eq. (25)) with $D_b$;
10. **End while**
11. **Return** the parameter set $\Theta$;

---

# 5. Experiment Setups and Results
## 5.1. Datasets

Two publicly-available LBSN datasets [5] (i.e., Gowalla and Brightkite) are used for our evaluation in this study. In the two datasets, a check-in record consists of user ID, check-in timestamp, POI ID, and the corresponding location. First of all, we preprocessed the two datasets to filter out inactive users who have fewer than 20 check-in records and unpopular POIs that have been visited for less than 20 times [15]. Secondly, we constructed check-in trajectories for each user. According to the definition of user trajectory, we split the check-in sequence of each user into trajectories of different lengths. As with the work of Cheng *et al.* [1], the interval threshold for any two successive check-ins was set to six hours. In other words, if the time interval between two consecutive check-ins is more than six hours, the two check-ins belong to two different trajectories. To alleviate the cold-start problem of next POI recommendation, we then removed those users with fewer than five trajectories from the two datasets. Next, we built two separate friend networks (i.e., LBSNs) composed of the remaining users and their friendships in the two datasets. Finally, we obtained two experimental datasets derived from the original datasets. Table 2 shows the statistics of the two experimental datasets.

TABLE 2. Statistics of the experimental datasets.

| Dataset | #Users | #Check-ins | #POIs | #Friendships | #Trajectories |
|---|---|---|---|---|---|
| Gowalla | 1,947 | 569,651 | 25,322 | 10,274 | 231,192 |
| Brightkite | 2,987 | 1,939,499 | 14,259 | 17,808 | 876,755 |

## 5.2. Baseline Approaches

To validate the effectiveness of the DAN-SNR in the next POI recommendation task, we compare it with the following seven competitive approaches.

(1) *FPMC-LR* (short for factorized personalized Markov chain for localized regions) [8]. It is a matrix factorization method that uses a Markov chain model to model customized sequential transitions of users. As an extension of FPMC [45], this method embeds the personalized Markov chain in check-in sequences and the localized regions in users' movement constraint.

(2) *PRME-G* (short for personalized ranking metric embedding with geographical influence) [9]. It is a metric embedding approach that models users' personalized check-in sequences by embedding users and POIs into a shared latent space. As an extension of PRME, this method utilizes a simple weighting scheme to fuse geographical influence.

(3) *ST-RNN* (short for spatial-temporal recurrent neural networks) [11]. It is an RNN-based model that incorporates spatial-temporal contexts in a recurrent architecture. This method extends an RNN and can model both temporal influence and geographical influence in each layer with specific transition matrices.

(4) *GRU4Rec+ST* (short for a gated recurrent unit for recommendations with spatial and temporal contexts) [46]. GRU4Rec is a session-based recommendation model that adopts an RNN-based framework. However, it is not designed for the next-POI recommendation task. In this study, we extend the GRU4Rec by embedding the spatial and temporal contexts of user check-ins into a compact vector representation.

(5) *SASRec+ST* (short for self-attention based sequential recommendation model with spatial and temporal contexts) [47]. SASRec is a sequential recommendation model that utilizes the self-attention mechanism. To be used in the next POI recommendation scenarios, we also extend the SASRec by embedding the spatial and temporal contexts of user check-ins into a compact vector representation.

(6) *ATST-LSTM* (short for an attention-based spatiotemporal long and short-term memory) [15]. It is an RNN-based model that leverages the attention mechanism. This method can model both temporal influence and geographical influence in each step and pay more attention to those relevant historical check-in records in a check-in sequence.

(7) *DGRec+ST* (short for a dynamic-graph-attention neural network for recommendations) [48]. DGRec is a session-based social recommender system that can capture social influence with a dynamic graph attention neural network. To compare with other approaches in the next POI scenarios, we incorporate the spatial and temporal information of check-ins in the same way.

Table 3 summarizes the eight approaches used in this study. Generally speaking, they are classified into three categories of commonly-used methods. First, the sequential POI recommendation approach using Markov chains (such as the FPMC-LR), embedding learning (such as the PRME-G), and neural networks (such as the ST-RNN and GRU4Rec+ST). Second, the attention-based POI recommendation approach, such as the SASRec+ST and ATST-LSTM. Third, the hybrid approach that fuses sequential influence and social influence, such as the DGRec+ST and DAN-SNR.

TABLE 3. Summary of the eight approaches used in this study.

| Feature | FPMC-LR | PRME-G | ST-RNN | GRU4Rec+ST | SASRec+ST | ATST-LSTM | DGRec+ST | DAN-SNR |
|---|---|---|---|---|---|---|---|---|
| SE | ✓ | ✓ | ✓ | ✓ | ✓ | ✓ | ✓ | ✓ |
| SP | ✓ | ✓ | ✓ | ✓ | ✓ | ✓ | ✓ | ✓ |
| TE | × | × | ✓ | ✓ | ✓ | ✓ | ✓ | ✓ |
| SO | × | × | × | × | × | × | ✓ | ✓ |
| AT | × | × | × | × | ✓ | ✓ | ✓ | ✓ |

SE, SP, TE, SO, and AT denote whether the given approach considers the sequential information, spatial information, temporal information, social information, and attention mechanism, respectively.

### 5.3. Evaluation Metrics

We evaluate the recommendation performance of all the eight approaches regarding two commonly-used metrics: Recall@$k$ and normalized discounted cumulative gain@$k$ (NDCG@$k$), where $k$ equals to five or ten. Note that we do not choose Precision@$k$ and F1-score@$k$ as primary evaluation measures. The main reasons are two-fold. First, $P@k$ (short for Precision@$k$) has a strong positive correlation with $R@k$ (short for Recall@$k$). Second, $R@k$ is more useful than $P@k$ to show a recommendation approach's capability of searching out more candidate POIs in the next POI recommendation scenarios.

$R@k$ measures how many of actual POIs in the test set are hit by the top-$k$ recommended items, formally defined as

$$R@k = \frac{1}{N}\sum_{u=1}^{N} R_u@k = \frac{1}{N}\sum_{u=1}^{N} \frac{|S_u(k)\cap V_u|}{|V_u|}, \quad (26)$$

where $S_u(k)$ denotes a set of the top-$k$ POIs recommended to user $u$, and $V_u$ means a collection of POIs that the user visits at the next moment in the test set. Note that $|V_u| = 1$.

$NDCG@k$ evaluates the ranking performance of a recommendation approach by considering the positions of actually visited POIs, formally defined as

$$NDCG@k = \frac{1}{N}\sum_{u=1}^{N}\frac{1}{Z_u}\sum_{j=1}^{k}\frac{2^{I(|\{s_u^j\}\cap V_u|)}-1}{\log_2(j+1)}, \quad (27)$$

where $I(\cdot)$ is an indicator function, $s_u^j$ is the $j$th recommended item in $S_u(K)$, and $Z$ is a normalization constant that is the maximum value of $DCG@k$.

Besides, we evaluate the recommendation efficiency of neural-network-based and attention-based approaches in terms of running time per batch.

### 5.4. Settings

We carried out our experiment on a Lenovo ThinkStation P910 Workstation with dual processors (2 x Intel Xeon E5-2660 v4, 2.0 GHz) and one graphics processing unit (GPU, NVIDIA TITAN X Pascal, 12GB). The operating system of the workstation was Microsoft Windows 10 (64-bit). All the program code used in our experiment was written in Python 3.7, and the deep learning framework we employed was TensorFlow[6] 1.2.0.

We built an L2L graph for each of the two datasets. However, the number of edges in the L2L graph grew exponentially with the increase of the number of nodes, which degraded the efficiency of graph embedding. To address this problem, we used an approximate solution to construct new L2L graphs according to the distance effect of human mobility [49]. For each POI, we picked out a certain number of POIs that have the shortest distance from the POI to rebuild a new L2L graph. The number of selected POIs was set to 20, which was far smaller than the number of POIs in the original L2L graphs. This solution facilitated the graph embedding process on L2L graphs.

The dimension size of five types of feature embeddings was set to 256. Feature embeddings were concatenated as the initial representation of a check-in behavior in a fully-connected layer. The number of neurons in the fully-connected layer was therefore set to 256. The lengths of the LTSC and STC were set to 200 and 50, respectively. If the number of user check-ins is smaller than the length of a channel, we conducted the operation of padding; otherwise, we chose only the latest 200 or 50 check-ins for the corresponding channel. In the two channels of the DAN-SNR, the number of the identical nonlinear layers was set to six; moreover, the number of attention heads in the self-attention layers was set to eight.

For the two datasets, we apportioned the data of user trajectories sorted in chronological order into training and test sets, with an 80-20 split. More specifically, the top 80% of trajectories were used as the training set, and the remainder of trajectories were used as the test set. For each target POI, the number of negative samples was set to 500. The batch size was set to 50. We employed Adam [44] as the optimizer and applied exponential decay in which the learning rate started at 0.001, and the decay rate was set to 0.96.

For more details of the settings of the proposed approach, please refer to the source code publicly available for download at https://github.com/drhuangliwei/DAN-SNR.

## 5.5. Results

### 5.5.1. *Recommendation performance*

TABLE 4. Comparison of different methods in recommendation performance.

| Metrics | Gowalla | | | | Brightkite | | | |
|---|---|---|---|---|---|---|---|---|
| Methods | R@5 | NDGG@5 | R@10 | NDGG@10 | R@5 | NDGG@5 | R@10 | NDGG@10 |
| FPMC-LR | 0.0543 | 0.1133 | 0.1297 | 0.1195 | 0.1197 | 0.1267 | 0.1407 | 0.1506 |
| PRME-G | 0.0769 | 0.1276 | 0.1481 | 0.1317 | 0.1245 | 0.1331 | 0.1612 | 0.1708 |
| ST-RNN | 0.0904 | 0.1282 | 0.1645 | 0.1648 | 0.1736 | 0.1632 | 0.1913 | 0.1854 |
| GRU4Rec+ST | 0.1004 | 0.1431 | 0.1885 | 0.1652 | 0.1853 | 0.1845 | 0.2214 | 0.2145 |
| SASRec+ST | 0.1227 | 0.1502 | 0.1964 | 0.1858 | 0.1934 | 0.1851 | 0.2415 | 0.2234 |
| ATST-LSTM | 0.1336 | 0.1537 | 0.1961 | 0.1898 | 0.1965 | 0.1863 | 0.2598 | 0.2328 |
| DGRec+ST | 0.1576 | 0.1599 | 0.2214 | 0.2057 | 0.2045 | 0.1933 | 0.2811 | 0.2608 |
| DAN-SNR | **0.1832** | **0.1783** | **0.2554** | **0.2219** | **0.2332** | **0.2214** | **0.3052** | **0.2823** |

Table 4 presents a comparison of the eight approaches in recommendation performance on the two datasets. The numbers shown in bold represent the best result of each column in Table 4. In general, the more the information, the better the approach. Because the PFMC-LR and PRME-G only use sequential and spatial data, they performed the worst among the eight methods regarding the two evaluation metrics.

---

[6] https://www.tensorflow.org/

Besides, the FPMC-LR employs the Markov chain method to model sequential patterns, implying that it cannot capture the long-term sequential influence. Compared with the PFMC-LR and PRME-G, the ST-RNN and GRU4Rec+ST made improvements in recommendation performance because they consider temporal and geographical influence further. Although the SASRec+ST and ATST-LSTM incorporate the same information used by the ST-RNN and GRU4Rec+ST, they achieved better results on the two datasets. The results indicate that leveraging the attention mechanism to model sequential influence can indeed improve recommendation performance. The DGRec+ST and DAN-SNR take into account four types of information and leverage the attention mechanism. As a result, they achieved the best results on both the two datasets.

Compared with the state-of-the-art DGRec+ST, the $R@5$, $NDGG@5$, $R@10$, and $NDGG@10$ values of the DAN-SNR were increased by 16.24%, 11.51%, 15.36%, and 7.88%, respectively, on the Gowalla dataset. For the Brightkite dataset, the performance improvements regarding the above four evaluation metrics were 14.03%, 14.54%, 8.57%, and 8.24%, respectively. The primary reasons that contribute to the state-of-the-art results are three-fold. First, the proposed approach makes full use of all four types of information available. Second, the DAN-SNR uses the self-attention mechanism to model user preference, and the results indicate that it is more effective than the Markov chain model (such as the PFMC-LR) and RNN-based neural networks (such as the ST-RNN, GRU4Rec+ST, ATST-LSTM, and DGRec+ST). Third, although the DGRec+ST also incorporates social information, it only models the short-term preferences of users within a session using an RNN architecture. Instead, our approach considers more historical check-ins of each user and the user's friends in a unified framework and measure the behavioral correlations between different check-ins adaptively using the self-attention mechanism.

5.5.2. *Recommendation efficiency*

We then conducted an efficiency analysis on six neural-network-based and attention-based methods, i.e., the ST-RNN, GRU4Rec+ST, SASRec+ST, ATST-LSTM, DGRec+ST, and DAN-SNR. To compare them in the same settings, the batch size, the embedding dimension, and the trajectory length of the other five approaches were set to 50, 256, and 50, respectively. For each of the two datasets, we calculated the time to run a batch under the same experimental environment on the whole dataset.

Table 5 shows a comparison of the six approaches in recommendation efficiency on the two datasets. The numbers shown in bold represent the best result of each row in Table 5, and #Params denotes the number of parameters involved in an approach. It is evident from Table 5 that the SASRec+ST and DAN-SNR, which leverage only the self-attention mechanism, work faster than the other four RNN-based methods, namely ST-RNN, GRU4Rec+ST, ATST-LSTM, and DGRec+ST. The SASRec+ST runs a batch with the minimum amount of time, followed by the DAN-SNR that has the maximum number of parameters. Compared with the SASRec+ST, our approach achieved 18.7% and 19.6% improvements in $NDCG@5$ on the two datasets, respectively. This result indicates that the DAN-SNR can make a better trade-off between performance and efficiency than the other five approaches.

TABLE 5. Comparison of different methods in recommendation efficiency.

| Dataset | Metric | ST-RNN | GRU4Rec+ST | SASRec+ST | ATST-LSTM | DGRec+ST | DAN-SNR |
|---|---|---|---|---|---|---|---|
| | #Params | 5,447,680 | 14,083,306 | **2,242,560** | 10,383,800 | 6,814,976 | 18,650,112 |
| Gowalla | Time(s)/batch | 0.52 | 3.62 | **0.15** | 2.24 | 2.03 | 0.25 |
| | *NDGG@5* | 0.1282 | 0.1431 | 0.1502 | 0.1537 | 0.1599 | **0.1783** |
| Brightkite | Time(s)/batch | 0.54 | 3.75 | **0.17** | 2.38 | 2.36 | 0.28 |
| | *NDGG@5* | 0.1632 | 0.1845 | 0.1851 | 0.1863 | 0.1933 | **0.2214** |

## 5.6. Attention Visualization

### 5.6.1. *Self-attention visualization*

Since we utilize the self-attention mechanism to model the dependencies between any two historical check-ins, in this subsection, we visualize such dependencies in the self-attention layer of the STC. Fig. 3 illustrates a qualitative analysis of a Gowalla user's 12 check-in records in three days. In Fig. 3, the left part plots the user's trajectory with Bing Maps[7], and the right part depicts the correlations between check-ins in the self-attention layer of the fifth and sixth nonlinear layers, which correspond to the lower and upper rows, respectively. For each check-in of the user's trajectory, we display the check-in time and POI category. The self-attention weights of all the check-ins in this trajectory are visualized with correlation lines and their color intensity. The darker the line, the higher the attention weight.

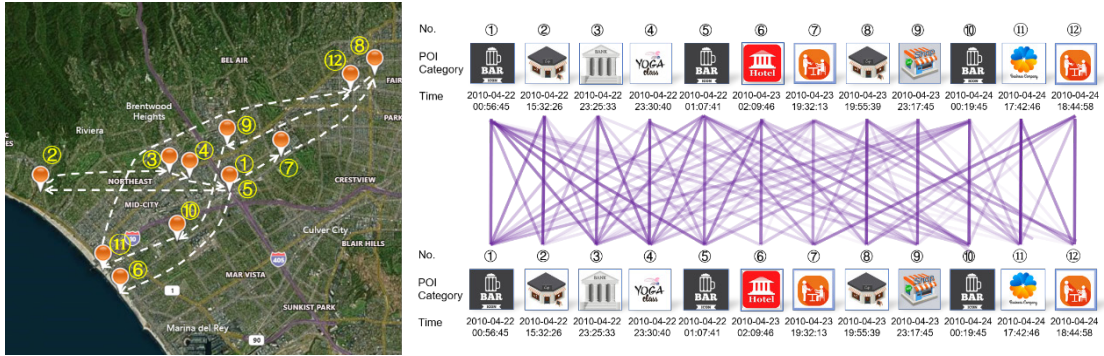

Fig. 3. An example of visualizing the dependencies between historical check-ins via the self-attention mechanism.

We find some interesting behavioral patterns of the user from Fig. 3. Firstly, a check-in behavior at one moment correlates highly with a small number of historical check-ins. For example, the sixth check-in has no significant correlations with previous ones except the fourth and fifth check-ins. This finding reflects the randomness and uncertainty of individual behavior. Secondly, long-range dependencies exist among the user's check-ins. For example, although there is a long distance between the tenth and first check-ins, there is a relatively high correlation between them, mainly due to the periodicity of the user's daily habit. In brief, the results mentioned above suggest that the DAN-SNR can indeed model users' sequential patterns better via the self-attention mechanism.

### 5.6.2. *Social influence visualization*

Because the DAN-SNR weighs the contribution of check-ins of a user's friends by the self-attention mechanism, in this subsection, we picture the effect of social influence on individual behavior over time. Fig. 4 displays the impact of a randomly-selected Gowalla user's friends on the user's current check-in behavior across eight consecutive timestamps. In this heat map, the X-axis represents the eight successive check-ins of the user, the Y-axis represents the user's nine friends, and each cell $(i, j)$ denotes the impact of the $j$th friend on the $i$th check-in behavior. Note that we use the attention weights from the Vanilla attention layer of the LTSC to express the effect of a friend's historical check-in on the user's current movement. Furthermore, we represent the total impact of a friend by averaging the attention weights of all historical check-ins of the friend.

Fig. 4 shows that the social influence of the user's friends varies from person to person. For example, in the sixth column, the sixth friend of the user has a significant impact on this check-in, which does not appear to be affected by the third friend. Besides, the influence of the same friend on the user's movement changes over time. For example, cells (5, 6) and (6, 6) look nearly opposite: one very white and one very dark, suggesting that the sixth friend has the opposite effect on the two successive check-ins. Moreover, this result indicates that the social influence of users is dynamic and context-dependent.

---

[7] https://www.microsoft.com/en-us/maps

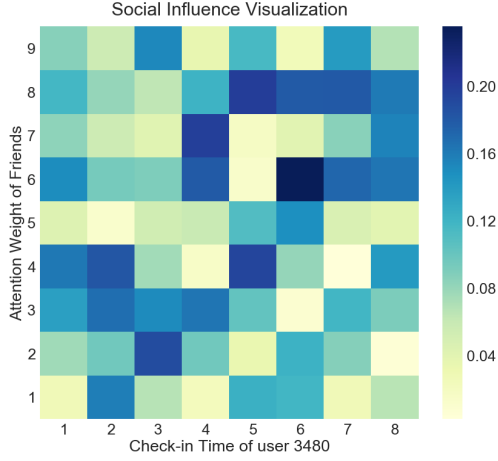

Fig. 4. An example of visualizing the social influence of friends on a user's movement over time.

## 6. Discussion
### 6.1. Sensitive Analysis of Parameters
#### 6.1.1. *Number of embedding dimensions*

The number of embedding dimensions is essential to the DAN-SNR. The higher this parameter, the stronger the representation ability of our approach. However, high values of this parameter may lead to the overfitting problem. Fig. 5 presents the effect of this parameter on the recommendation performance of the DAN-SNR on the Gowalla and Brightkite datasets. When the number of embedding dimensions is below 200, the $R@10$ and $NDCG@10$ values significantly increase with the increase of this parameter. This result implies that the representation ability of our approach has grown remarkably. As the number of embedding dimensions exceeds 256, there is a slight decline in recommendation performance in terms of $R@10$ and $NDCG@10$, suggesting that the representation ability of the DAN-SNR has reached its limit. Therefore, the number of embedding dimensions was set to 256 in this study.

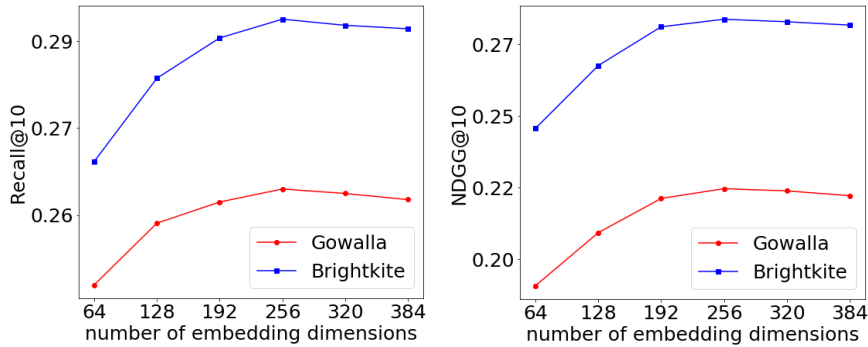

Fig. 5. Performance tuning with different embedding dimensions.

#### 6.1.2. *Number of negative samples*

The number of negative samples is another critical parameter of the DAN-SNR. The performance of recommendation models will get improved with the increase of this parameter, which has been proved by some previous studies on representation learning [15],[50]. However, if all unobserved data is used as negative examples, the computational complexity of a model will increase substantially. Fig. 6 shows the effect of this parameter on the recommendation performance of the DAN-SNR on the two datasets. Note that "All" in Fig. 6 denotes a specific case that all unobserved POIs were selected to be negative samples without data sampling. As the number of negative samples increases, the performance of our approach tends to improve considerably. However, the *R@10* and *NDCG@10* values remain unchanged after this parameter exceeds 500. Moreover, there is a visible decrease in the performance of the DAN-SNR when this parameter is higher than 1,000. Therefore, the number of negative samples was set to 500 in our experiment.

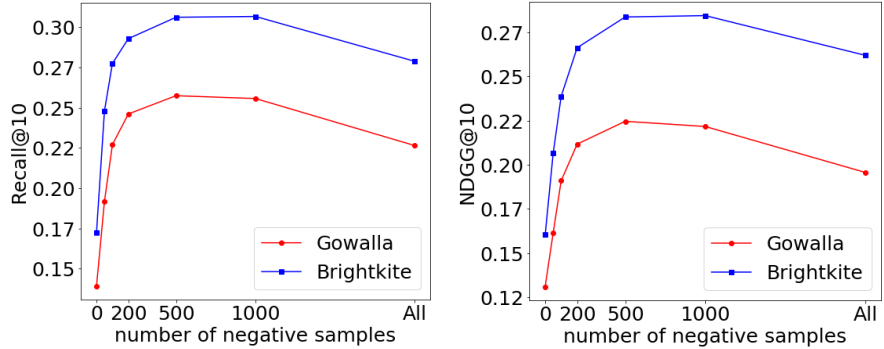

Fig. 6. Performance tuning with different negative samples.

## 6.2. Ablation Study

### 6.2.1. *Personal preference versus social influence*

For each target user, the DAN-SNR generates his/her final representation that combines the user's past behavior and context-dependent social influence. To see how the above two "features" affect our method's performance, we designed two DAN-SNR variants: DAN-SNR-self and DAN-SNR-social. The former removed all check-ins of each user's friends and leveraged only the user's historical check-ins. The latter took into account only social influence while disregarding users' personal preferences mined from their past behavior. The difference between the two variants is on what channel they used. DAN-SNR-self employs two channels to model long- and short-term user preferences, respectively, while DAN-SNR-social uses one channel to model social influence.

Table 6 presents the recommendation performance of our approach and its two variants on the two datasets. As shown in Table 6, DAN-SNR-self outperforms DAN-SNR-social across the two datasets, suggesting that a user's personal preference contributes more to the user's next move than his/her friends' influence. Therefore, modeling user preference based on historical check-ins is a necessary part of our method. Once the DAN-SNR makes use of both the two "features," there is a marked increase in the values of the two evaluation metrics, which is, of course, the primary motivation of this study. As a result, it is crucial to model user preference and social influence together in the next POI recommendation task to improve recommendation performance further.

TABLE 6. Comparison between our approach and its variants in recommendation performance.

| Metrics | Gowalla | | | | Brightkite | | | |
| Methods | R@5 | NDGG@5 | R@10 | NDGG@10 | R@5 | NDGG@5 | R@10 | NDGG@10 |
| --- | --- | --- | --- | --- | --- | --- | --- | --- |
| DAN-SNR-self | 0.1540 | 0.1413 | 0.2123 | 0.1931 | 0.1897 | 0.1784 | 0.2674 | 0.2364 |
| DAN-SNR-social | 0.1090 | 0.0960 | 0.1496 | 0.1573 | 0.1442 | 0.1204 | 0.1612 | 0.1772 |
| DAN-SNR | 0.1832 | 0.1783 | 0.2554 | 0.2219 | 0.2332 | 0.2214 | 0.3052 | 0.2823 |

### 6.2.2. *Short-term preference versus long-term preference*

Since user preference plays an essential role in personalized POI recommendation, the DAN-SNR provides a mechanism to encode users' short- and long-term preferences. Then, we studied the impact of two different types of user preferences on recommendation performance and also designed two variants: DAN-SNR-long and DAN-SNR-short. The former took into consideration long-term user preference, while the latter leveraged short-term user preference. Note that we deleted the "feature" of social influence from both the two variants.

Fig. 7 displays a comparison between our approach and its variants in performance on the Gowalla and Brightkite datasets. DAN-SNR-short achieved, unsurprisingly, higher *R@*10 and *NDCG@*10 values than DAN-SNR-long across the two datasets. The main reason is that short-term preference can capture users' changing needs in context-dependent scenarios better. Therefore, short-term user preference has a more significant impact on recommendation performance. However, it is worth noting that DAN-SNR-short offered a slight performance advantage over DAN-SNR-long. Compared with the former, DAN-SNR-long analyzed each user's current check-in trajectory and did not extract sequential patterns from these trajectories. In other words, long-term user preference captures long-range dependencies between historical check-ins. When our method took into account long- and short-term preferences together, its performance was improved significantly.

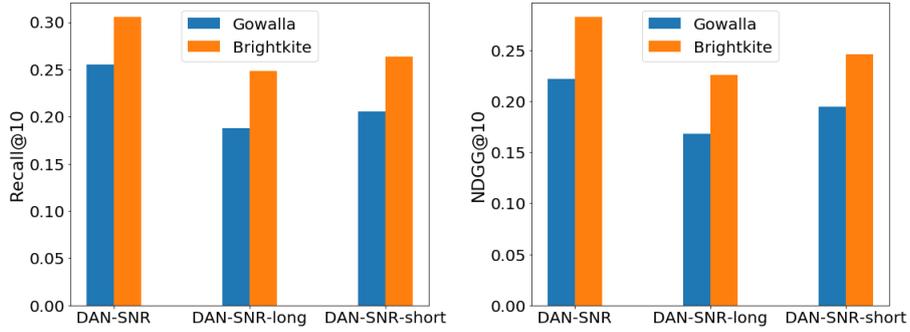

Fig. 7. Effect of two different types of user preferences on recommendation performance.

## 7. Conclusion

Next (or successive) POI recommendation is a challenging task of POI recommendation and has been studied in recent years. In this study, we discuss a new research topic, i.e., social-aware next POI recommendation. By incorporating sequential influence, temporal influence, spatial influence, and social influence, we design and implement a deep attentive neural network called DAN-SNR. More specifically, the DAN-SNR can model the context-dependent social influence by capturing the behavioral correlation between the target user and his/her friends. By leveraging the self-attention mechanism rather than using the RNN architecture, it can better model long-range dependencies between historical check-ins of each user regardless of the distance between them. Besides, experimental results on two public LBSN datasets indicate that the proposed approach outperforms seven competitive baselines regarding two commonly-used metrics, i.e., *Recall* and *NDCG*.

It is worth noting that our work may play a promising role in many application scenarios, such as location-based item recommendation, mobile advertising, and travel assistant. In particular, a smart travel assistant can create personalized user profiles by automatically learning historical check-in records of a user and the mobility behaviors of the user's friends, as well as help to recommend possible POIs to go next in real-time. In the future, we will incorporate more context information, such as visual and text information associated with users and POIs, into the DAN-SNR to improve performance further.

## Acknowledgment

This work was partially supported by the National Key Research and Development Program of China (No. 2018YFB1003801), National Science Foundation of China (Nos. 61972292 and 61672387). Yutao Ma is the corresponding author of this paper.